\documentclass{moriond}

\def\Journal#1#2#3#4{{#1} {\bf #2}, #3 (#4)}

\def\AMP{\em American Journal of Physics}
\def\AA{\em Astronomy \& Astrophysics}
\def\AJ{\em The Astronomical Journal}
\def\APJ{\em The Astrophysical Journal}
\def\APJL{\em The Astrophysical Journal Letters}
\def\CQG{\em Class. Quant. Gravity}
\def\MNRAS{\em Month. Not. Roy. Ast. Soc.}
\def\NAT{\em Nature}

\def\PRL{\em Phys. Rev. Lett.}
\def\PRD{{\em Phys. Rev.} D}
\def\PRX{{\em Phys. Rev.} X}
\def\PR{\em Physical Review}
\def\SCI{\em Science}

\def\be{\begin{equation}}
\def\ee{\end{equation}}
\def\bea{\begin{eqnarray}}
\def\eea{\end{eqnarray}}

\newcommand{\Photo}{\includegraphics[height=35mm]{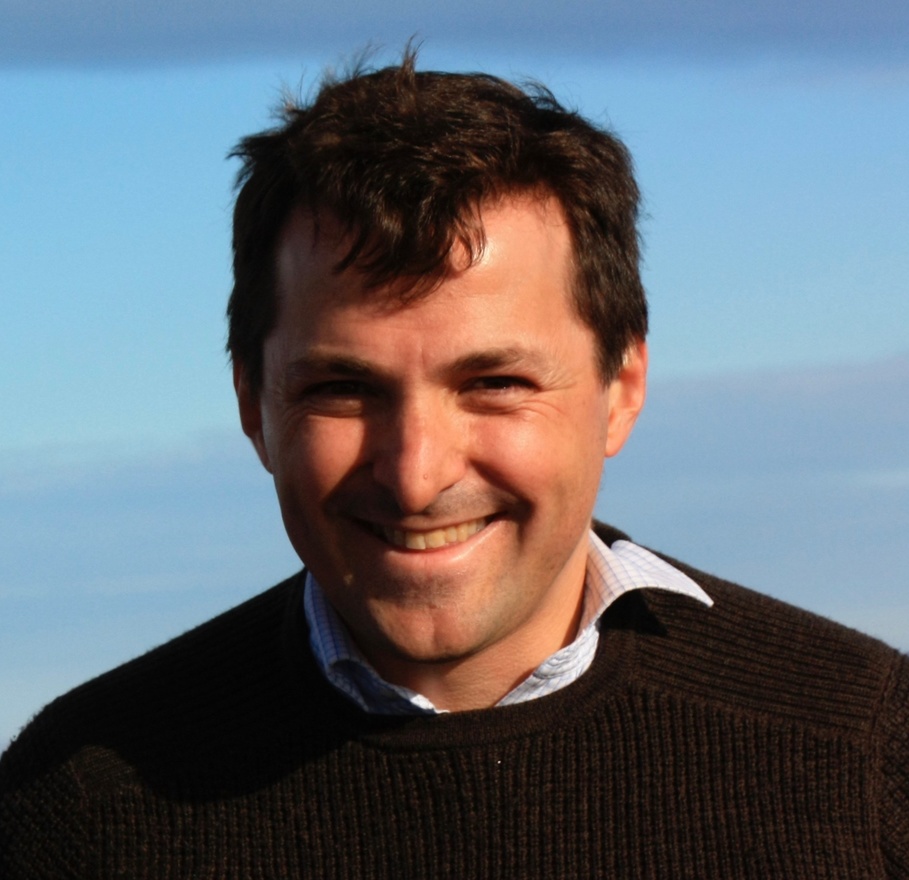}}

\begin{document}
\vspace*{4cm}
\title{Tests of gravity theories with pulsar timing}

\author{ Paulo C. C. Freire }

\address{Max-Planck-Institut f\"ur Radioastronomie, 69 Auf dem H\"ugel,\\
D-53121, Bonn, Germany}

\maketitle\abstracts{
Over the last few years, a set of new results from pulsar timing has introduced much tighter constraints on violations of the strong equivalence principle (SEP), either via a direct verification of the universality of free fall for a pulsar in a triple star system, or from tests of the nature of gravitational waves, in particular a search for dipolar gravitational wave emission in a variety of binary pulsars with different masses. No deviations from the SEP have been detected in our experiments. These results introduce some of the most stringent constraints on several classes of alternative theories of gravity and complement recent results from the ground-based gravitational wave detectors.}

\section{Timing binary pulsars}

The 1974 discovery of the first binary pulsar (PSR B1913+16) by Russell Hulse and Joseph Taylor~\cite{ht75} offered us a powerful new tool for understanding gravity and spacetime. The system consists of two neutron stars (NSs), one of them a radio pulsar with a spin period of 59 ms, its orbital period is 7h 45m and eccentricity is 0.617. This high eccentrivity means that the longitude of periastron, $\omega$ - the point where the pulsar is closer to its companion - can be measured precisely. Subsequent timing of this system measured three relativistic effects on the orbit: the increase in $\omega$ with time, $\dot{\omega}$ (which is analogous to the relativistic perihelion shift in Mercury), the so-called ``Einstein delay'', a slow-down of the pulsar spin near periastron caused by special relativistic time dilation and gravitational redshift ($\gamma_{\rm E}$), and the decay of the orbit, caused by loss of orbital energy ($\dot{P}_{\rm b}$). Each effect depends on the two unknown masses of the system, the dependence is specified by a particular theory of gravity. The self-consistency of the masses obtained from the measured relativistic effects using the equations of general relativity (GR) means that this theory passed the test posed by the measurement of $\dot{\omega}$, $\gamma_{\rm E}$ and $\dot{P}_{\rm b}$ in the PSR~B1913+16 system. The loss of orbital energy associated with the latter parameter is caused by the emission of gravitational waves (GW); the fact that the measured $\dot{P}_{\rm b}$ agrees with GR's predictions for the PSR~B1913+16 system confirmed experimentally the existence of GWs~\cite{tw82,tw89,dt92,wh16}. This was a major impulse to the start of the era of GW astronomy.

\section{Recent results}

Currently, we know more than 3300 pulsars in our Galaxy, the associated globular cluster system and also in the Magellanic clouds~\cite{mht+05}. Of these, 10\% are in binary systems. Despite this, measurements of relativistic parameters are rare: only 40 of them have had the component masses measured to a relative precison better than 15\%, with an additional 17 systems where the total system mass has been measured via $\dot{\omega}$. Of all these binaries, 17 systems (including B1913+16) have been confirmed as double neutron star systems (DNSs) from strong stellar evolution arguments, with an additional eight candidates\footnote{For a complete, up-to-date lists on binary pulsar mass measurements and a list of pulsars in DNS systems see \url{https://www3.mpifr-bonn.mpg.de/staff/pfreire/NS_masses.html}.}.

\subsection{The Amazing ``Double Pulsar'' system}\label{subsec:double_pulsar}

One discovery, that of PSR~J0737$-$3039A~\cite{bdp+03}, was an extremely lucky find, for several reasons. First, the system happened to be, at the time, the most compact DNS known (orbital period of 2h 27m, $e = 0.0878$); thus all relativistic effects observed in PSR~B1913+16 ($\dot{\omega}$, $\gamma_{\rm E}$ and $\dot{P}_{\rm b}$) can also be measured very precisely in this system. Second, the system happens to have the highest inclination known for any binary pulsar, $89.35^\circ$! This means that further independent and precise mass constraints can be obtained from the Shapiro delay in the arrival times of the pulses. Third, the NS companion has also been detected as a radio pulsar, PSR~J0737$-$3039B; the measurement of the sizes of both orbits yields the NS mass ratio~\cite{lbk+04}.

This system now includes seven independent and precise strong-field tests of GR, which the theory passes with flying colours~\cite{ksm+21}. Some highlights:
\begin{itemize}
    \item The most precise test of the GR quadrupole formula for GW emission: the measured orbital decay ($-$39.3770 $\pm$ 0.0025 microseconds per year, the uncertainty is a 68.3\% confidence limit) agrees with the GR prediction within the tiny relative uncertainty of $6.3 \times 10^{-5}$. This is 25 times more precise than for the Hulse-Taylor binary!
    \item They also include detection, for the first time, the detection of the effect of light bending in the timing of the pulsar. This experiment probes the propagation of radiation in a spacetime that has $10^6$ times or more the curvature of any non-pulsar experiments.
    \item In addition, there are also important second-order effects on the periastron advance. One of these is the Lense-Thirring effect, the frame dragging caused by the rotation of pulsar A. The latter effect will, when measured more precisely, be important for measuring the moment of inertia of pulsar A.
\end{itemize}

 The precise measurement of the mass and moment of inertia of a neutron star will represent powerful constraints on the largely unknown properties of super-dense matter in the centres of neutron stars, in particular the equation of state of dense matter (EOS)~\cite{hkw+20,ksm+21}.

\subsection{Asymmetric binaries}

The strong equivalence principle (SEP) is the assertion that the outcome of any local experiment, {\em including any gravitational experiments}, in a freely falling laboratory is independent of the velocity of the laboratory and its location in spacetime. There are suggestive arguments that GR is the only theory in four spacetime dimensions that fully embodies the SEP~\cite{dls15,will18}, with all other viable alternatives violating it at some level.

Such a violation would have observable consequences, chief among these a violation of the universality of free fall (UFF). One of the resulting effects would be the emission of dipolar GWs (DGW)~\cite{ear75}. Detecting DGW emission would falsity GR, because in GR GWs are, to leading order, quadrupolar; not detecting it will necessarily constrain many alternative theories of gravity.

DGW emission would be observable in a binary pulsar as an orbital decay that is faster than that predicted by GR. 
Although the double pulsar system yields a very precise test of the GR quadrupole formula, it does not by itself exclude DGW emission. This happens for two reasons: first, the latter is, within certain gravity theories, enhanced by the {\em difference} in compactness of the two components in the system; for this reason, pulsar - WD systems can be of great value for this enterprise. Second, DGW emission can be greatly enhanced for NS masses larger than those of the components of the double pulsar system~\cite{de93}.

For these reasons, we have over the years conducted orbital decay experiments for a variety of binary systems: PSR~J1738+0333~\cite{fwe+12}, PSR~J0348+0432~\cite{afw+13}, PSR~J1909$-$3744~\cite{lgi+20}, PSR J1012+5307~\cite{ddf+20}, and PSR~J2222$-$0137~\cite{gfg+21}, all pulsar - WD systems. We have also been timing PSR~J1913+1102, an asymmetric DNS~\cite{ffp+20}. Their timing yields tight limits on DGW emission that are valid for the observered range of NS masses and are independent of the EOS.

\subsection{The pulsar in a triple star system}\label{subsec:triple}

In 2014, the first pulsar in a triple star system, PSR~J0337+1715, was discovered~\cite{rsa+14}. Such systems were expected to provide some of the most stringent tests of the UFF because, if NSs and WDs accelerate differently in the gravitational field of the outer WD, the inner PSR - WD pair in the system would show ``orbital polarization''~\cite{fkw12} known as the Nordtvedt effect~\cite{nor68}.

The suitability of PSR~J0337+1715 for this test was demonstrated in 2018, with the finding that NSs and WDs fall with the same acceleration within 2.6 parts in $10^6$~\cite{agh+18}. A more recent study, using a completely independent data set (from the Nan\c{c}ay telescope) and an improved timing analysis and theoretical treatment of the three-body problem in alternative theories of gravity has found a lower level of systematic noise, and that NSs and WDs fall at the same rate within 2 parts in $10^6$; this is statistical 95\% confidence limit~\cite{vcf+20}. An update on the timing of this system is provided by G. Voisin in these proceedings.

\section{Consequences of these experiments}

\begin{figure}
\begin{minipage}{0.4\linewidth}
\centerline{\includegraphics[width=\linewidth]{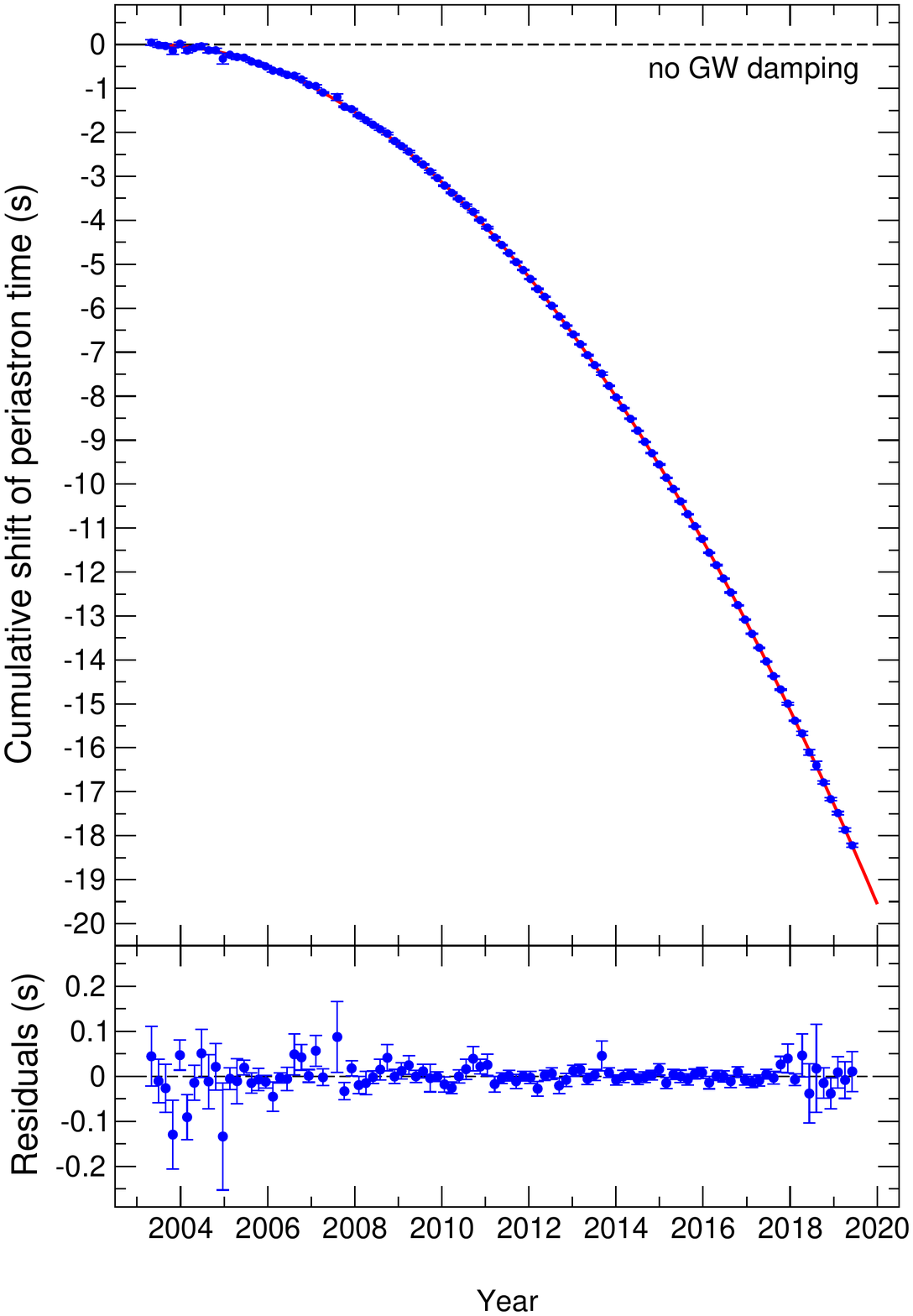}}
\end{minipage}
\hfill
\begin{minipage}{0.6\linewidth}
\centerline{\includegraphics[width=\linewidth]{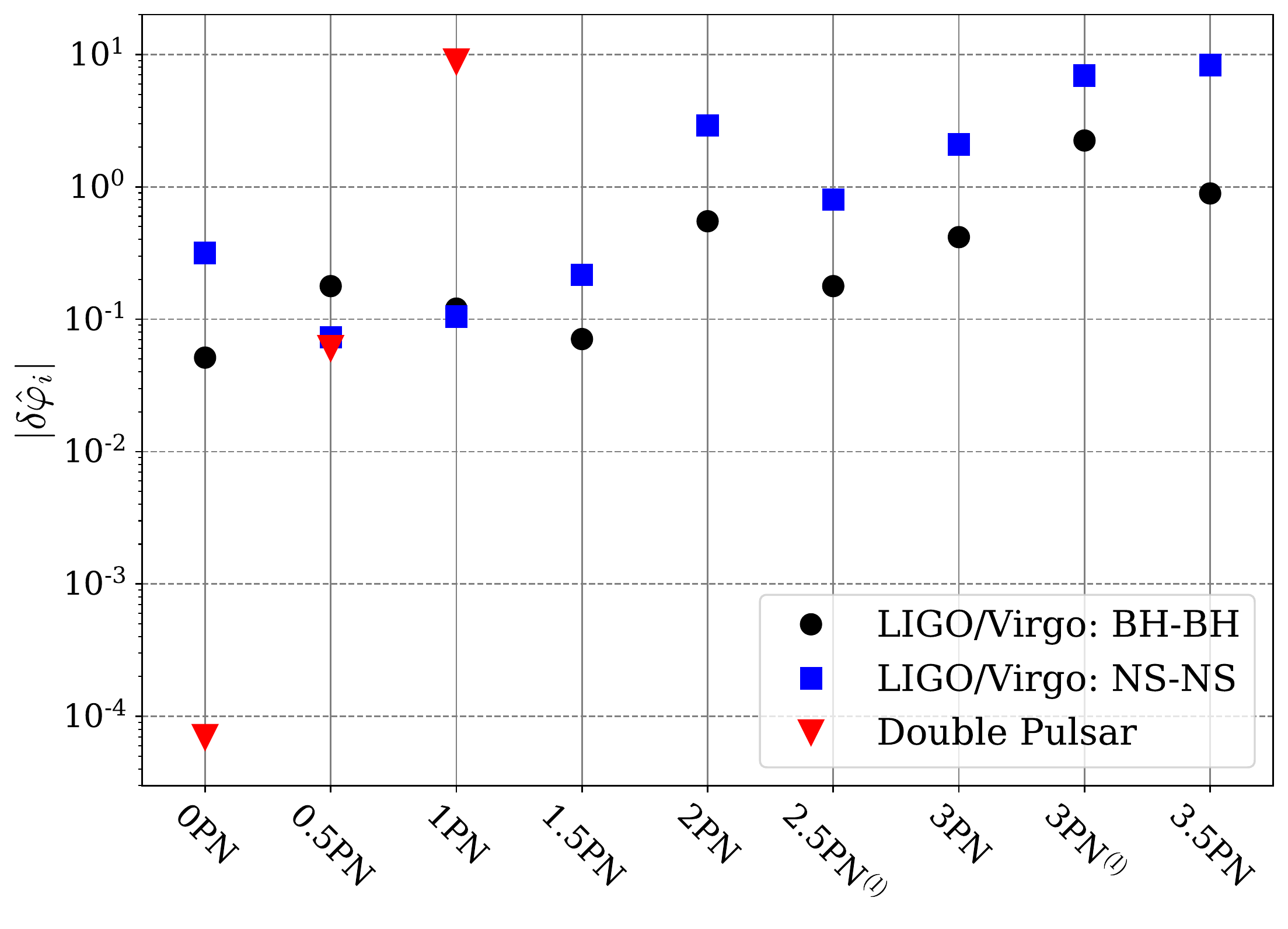}}
\end{minipage}
\caption[]{ {\em Left}: Comparison between the cumulative shift in periastron time predicted by GR (in red) and our measurements (in blue), which are consistent with the former within measurement uncertainty. {\em Right:} Limits on a set of parameters that quantify deviations from the GR prediction for GW emission at successive ``Post-Newtonian'' orders, derived from the double pulsar timing (red) and different LIGO-Virgo observations (blue and black). The results from the double pulsar are three orders of magnitude more constraining for the leading order term. From Kramer et al. (2021).}
\label{fig:orbital_decay}
\end{figure}

\begin{figure}
\begin{minipage}{0.57\linewidth}
\centerline{\includegraphics[width=\linewidth]{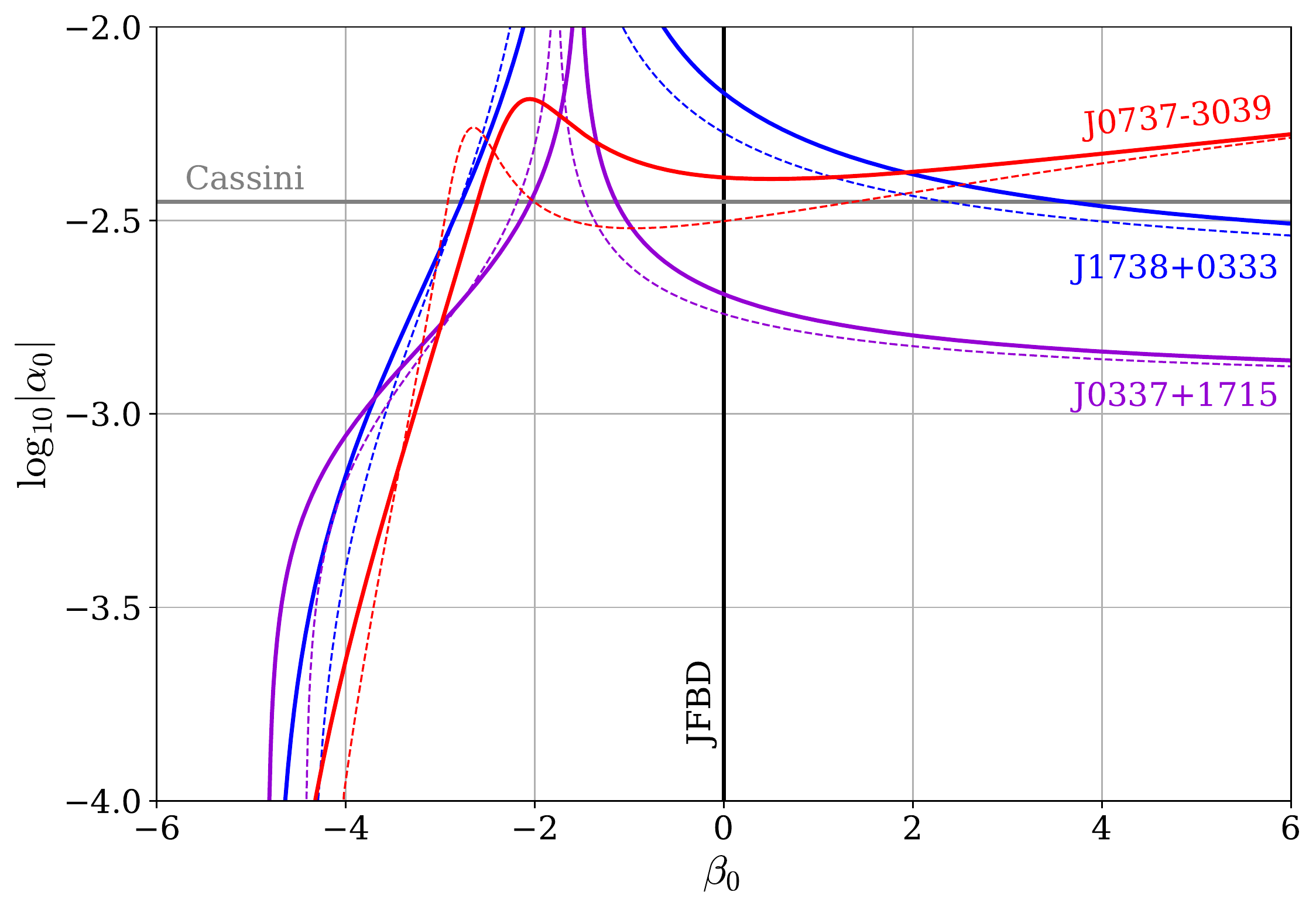}}
\end{minipage}
\hfill
\begin{minipage}{0.43\linewidth}
\centerline{\includegraphics[width=\linewidth]{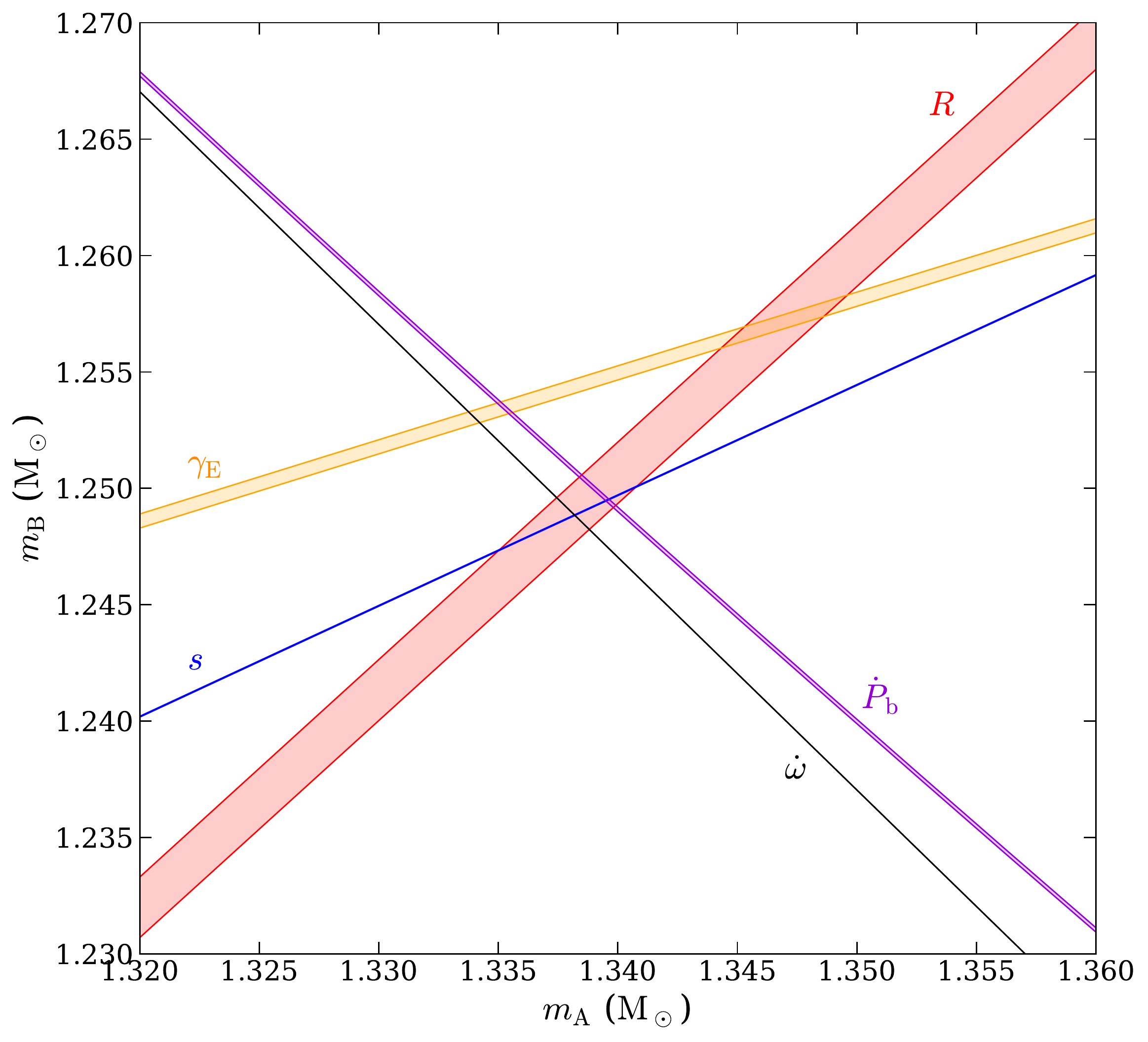}}
\end{minipage}
\caption[]{{\em Left}: Constraints on the Damour - Esposito-Far\`ese scalar-tensor theories of gravity (DEF gravity)~\cite{de93}. The double pulsar and triple system introduce the tightest limits in different regions of the DEF parameter space. In the right plot, we see a mass-mass diagram obtained with a member of Bekenstein's 2004 family of Scalar-Vector-Tensor theories of gravity. The theory cannot produce a pair of masses that explains all observed relativistic effects in the double pulsar system and at the same time produce the MOND phenomenon in the rotation of galaxies; the theory is therefore excluded. From Kramer et al. (2021).}
\label{fig:theories}
\end{figure}

\begin{figure}
\begin{minipage}{\linewidth}
\centerline{\includegraphics[width=\linewidth]{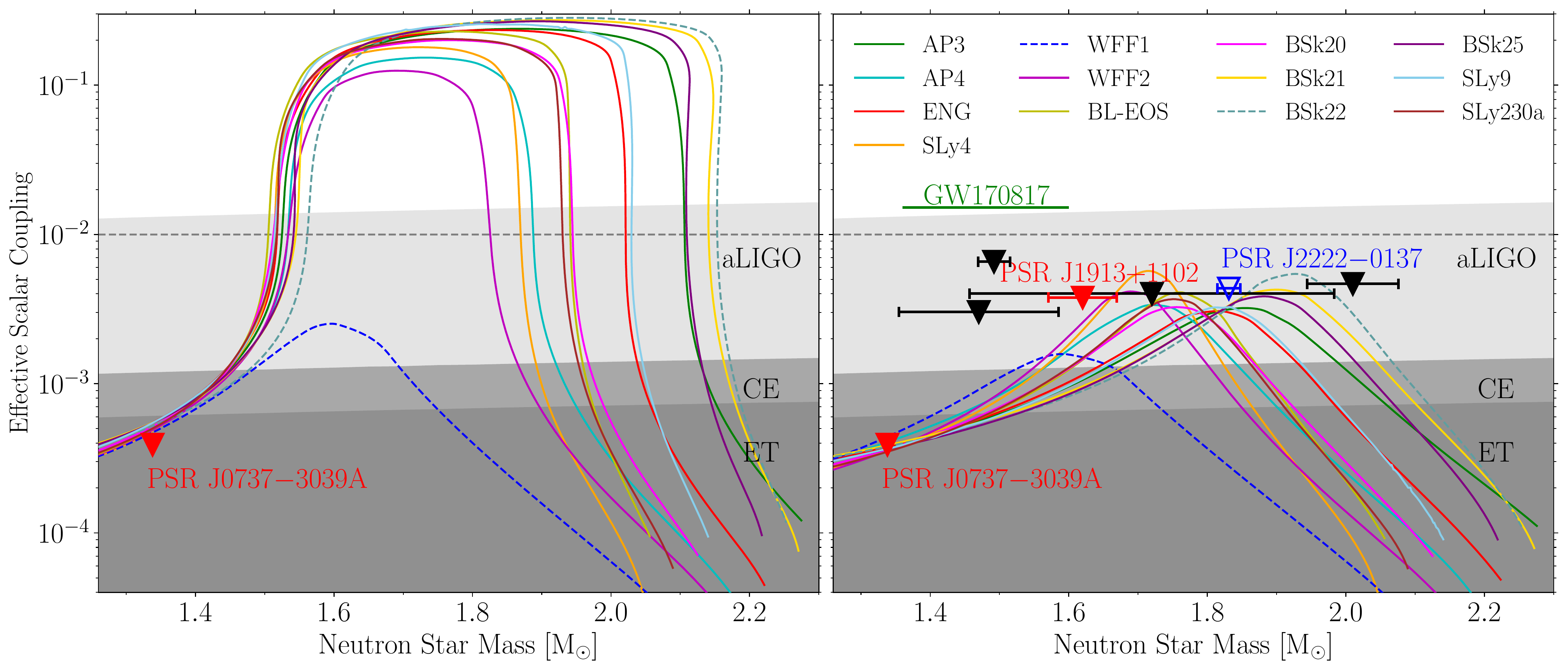}}
\end{minipage}
\caption[]{Limits on scalar coupling for different pulsars (triangles) and for DEF gravity, using different EOSs (coloured lines). The double pulsar does not constrain the possibility that more massive NSs might scalarise very significantly; this was excluded by timing systems with more massive NSs. From Zhao et al. (2022).
}
\label{fig:spontaneous_scalarization}
\end{figure}

In all experiments mentioned above, we have tested GR to very high precision, especially in the case of the double pulsar. This can be seen in Fig.~\ref{fig:orbital_decay}, where we see in the left panel the comparison between the cumulative shift in periastron time predicted by GR and our measurements, and in the right panel a comparison of constraints on deviations of GW emission from the GR prediction at successive PN orders\cite{aaa+16} for the double pulsar and LIGO/Virgo observations~\cite{ksm+21}. Clearly, these experiments provide complementary information, with the double pulsar experiment providing more constraining limits (by three orders of magnitude) for the leading order term!

We have not detected any of the consequences of UFF violation, either DGW emission or the Nordtvedt effect for the pulsar in the triple star system. 
As we can see in Fig.~\ref{fig:theories}, these non-detections result in tight constrains on several classes of alternative gravity theories~\cite{ksm+21}. 

Within some of these theories, there is a phenomenon called ``spontaneous scalarization'', where within a range of NS masses, the NS can acquire a very significant scalar ``charge''~\cite{de93}, greatly enhancing DGW emission. This phenomenon cannot be fully probed by the double pulsar test, because of the low mass of the NSs in that system (see left panel of Fig.~\ref{fig:spontaneous_scalarization}). Since DGW emission is not observed for any of the binaries we have been timing, which cover the full range of observed NS masses, we can put tight constraints on spontaneous scalarisation for all observed NS masses that are independent of the EOS of dense matter, at least for some specific classes of gravity theories~\cite{zfk+22} (see right panel in Fig.~\ref{fig:spontaneous_scalarization}).

\section*{Acknowledgments}

I thank the Max-Planck-Gesellschaft and the European Research Council for their support, and especially Michael Kramer and Norbert Wex for the wonderful collaboration over the years.

\end{document}